\documentclass[12pt,a4paper]{article}
\usepackage[utf8]{inputenc}
\usepackage[plainpages=false,pdfpagelabels=true]{hyperref}
\usepackage{amssymb,amsthm}
\usepackage[margin=1.5 in]{geometry}
\usepackage{a4wide}  
\usepackage{amsfonts}
\usepackage{amsmath}
\usepackage{bbm}
\usepackage{booktabs} 
\usepackage{ifpdf}

\ifpdf

\hypersetup{%
  pdftitle    = {RG-2 flow and  black hole entanglement entropy }
  pdfkeywords = {RG flow, RG-2 flow,entanglement entropy,black holes},
  pdfauthor   = {Oscar Lasso Andino},
  plainpages  = true,
  colorlinks  = true,
  citecolor   = blue,
  urlcolor    = blue,
  linkcolor   = black
}

\makeatletter
\@addtoreset{equation}{section}
\makeatother

\begin{document}

\vspace{1.5cm}

\begin{center}

{\Large {\bf RG-2 flow and  black hole entanglement entropy }}

\vspace{1.5cm}

\renewcommand{\thefootnote}{\alph{footnote}}
{Oscar Lasso Andino}\footnote{E-mail: {oscar.lasso [at] udla.edu.ec}}

\setcounter{footnote}{0}
\renewcommand{\thefootnote}{\arabic{footnote}}

\vspace{1.5cm}

{\it $^a$ Escuela de Ciencias Físicas y Matemáticas, Universidad de Las Américas,\\
C/. José Queri, C.P. 170504, Quito, Ecuador}\\ \vspace{0.3cm}

and \\
\vspace{0.3cm}

{\it Instituto de F\'{\i}sica Te\'orica UAM/CSIC\\
C/ Nicol\'as Cabrera, 13--15,  C.U.~Cantoblanco, E-28049 Madrid, Spain}\\ \vspace{0.3cm}

\vspace{1.8cm}

%%%%%%%%%%%%%%%%%%%%%%%%%%%%%%%%%%%%%%%%%%%%%%%%%%%%%%%%%%%%%%%%%%%%%%

{\bf Abstract}

\end{center}

\begin{quotation}
\noindent
We study the evolution of a Euclidean two dimensional black hole metric under the second loop renormalization group flow, the RG-2 flow. Since the black hole metric is non-compact (we consider it asymptotically flat) we adapt some proofs for the compact case to the asymptotically flat case. We found that the appearance of horizons during the evolution is related to the 
parabolicity condition of the flow. We also show that the entanglement entropy of the two dimensional Euclidean Schwarzschild black hole is monotonic under the RG-2 flow. We generalize the results obtained for the first loop approximation and discuss the implications for higher order loops 
\end{quotation}

\newpage
%%%%%%%%%%%%%%%%%%%%%%%%%%%%%%%%%%%%%%%%%%%%%%%%%%%%%%%%%%%%%%%%%%%%%%
%%%%%%%%%%%%%%%%%%%%%%%%%%%%%%%%%%%%%%%%%%%%%%%%%%%%%%%%%%%%%%%%%%%%%%
%%%%%%%%%%%%%%%%%%%%%%%%%%%%%%%%%%%%%%%%%%%%%%%%%%%%%%%%%%%%%%%%%%%%%%
%%%%%%%%%%%%%%%%%%%%%%%%%%%%%%%%%%%%%%%%%%%%%%%%%%%%%%%%%%%%%%%%%%%%%%
\pagestyle{plain}
%%%%%%%%%%%%%%%%%%%%%%%%%%%%%%%%%%%%%%%%%%%%%%%%%%%%%%%%%%%%%%%%%%%%%%
%%%%%%%%%%%%%%%%%%%%%%%%%%%%%%%%%%%%%%%%%%%%%%%%%%%%%%%%%%%%%%%%%%%%%%
%%%%%%%%%%%%%%%%%%%%%%%%%%%%%%%%%%%%%%%%%%%%%%%%%%%%%%%%%%%%%%%%%%%%%%
%%%%%%%%%%%%%%%%%%%%%%%%%%%%%%%%%%%%%%%%%%%%%%%%%%%%%%%%%%%%%%%%%%%%%%
%%%%%%%%%%%%%%%%%%%%%%%%%%%%%%%%%%%%%%%%%%%%%%%%%%%%%%%%%%%%%%%%%%%%%%

%\tableofcontents

\newpage

%%%%%%%%%%%%%%%%%%%%%%%%%%%%%%%%%%%%%%%%%%%%%%%%%%%%%%%%%%%%%%%%%%%%%%
%%%%%%%%%%%%%%%%%%%%%%%%%%%%%%%%%%%%%%%%%%%%%%%%%%%%%%%%%%%%%%%%%%%%%%
%%%%%%%%%%%%%%%%%%%%%%%%%%%%%%%%%%%%%%%%%%%%%%%%%%%%%%%%%%%%%%%%%%%%%%
%%%%%%%%%%%%%%%%%%%%%%%%%%%%%%%%%%%%%%%%%%%%%%%%%%%%%%%%%%%%%%%%%%%%%%

\section{Introduction}

The idea of irreversibility is very well known in thermodynamics, it leads to the concept of thermal entropy. In general, in a microscopic theory, such as a quantum system, there is not a preferred time direction. Its equations seem not to distinguish between any time direction. However, in a macroscopic system the irreversibility  arises naturally and it is stated as the second law of thermodynamics. If we want to understand the connection between, let's say thermodynamics and the microscopic phenomena, we should use a tool capable of providing a thoughtful analysis through different scales. Depending on the theory that we are working on, any irreversibility will be manifest as a monotonicity of a certain quantity under a given evolution parameter.\\
In quantum field theory, for example,  the renormalization group flow (RG flow) is directly related to irreversibility, and therefore in some sense, to entropy. Nevertheless, it has some particularities. There are some examples of microscopical irreversible processes along the RG flow, for instance, in \cite{Zamolodchikov:1986gt} it is  shown that there is a $c$-function monotonous under the 2-dimensional renormalization group flow. This function depends on the coupling constants and the renormalization scale $\Lambda$. Similar results have been proved in three and four dimensions \cite{Casini:2012ei,Komargodski:2011xv,Casini:2017vbe,Park:2018ebm,Casini:2016udt}. The irreversibility at quantum level can be related to entanglement entropy, this quantity is due to the quantum correlations between subsystems separated by a compact surface. Contrary to what happens in a macroscopic theory with the thermal entropy, this entanglement entropy is not a consequence of the lack of information about the microscopic states but it comes  from a fundamental property in quantum physics: entanglement.\\
Due to the fact that a black hole horizon divides the spacetime into two subsystems, in such a  way that an observer located in the zone outside the horizon  does not have access to the information inside the horizon,  the entanglement entropy of a black hole can, in principle, be calculated. And, as in the previous examples, we expect some type of monotonicity under some RG flow.\\
It is known \cite{Solodukhin:2006xv} that for a given conformal field theory with central charge $c$, the entanglement entropy of a two dimensional euclidean Schwarzschild black hole\footnote{In the gravity dual the holographic entanglement entropy follows an area law, which resembles us the Bekenstein-Hawking entropy of the black hole theory. In black holes living in the boundary of an AdS space there is an holographic interpretation for their entanglement entropy.} is given by\footnote{Here $h(x)$ is the metric function, $\epsilon$ is the UV regulator, $\beta_{H}$ is the Hawking temperature and $L$ is the size of the box where we have put the black hole}
\begin{equation*}
S=\frac{c}{12}\int_{x_{h}}^{L}\frac{dx}{h(x)}\left(\frac{4\pi}{\beta_{H}}-h'\right)+\frac{c}{6}\ln\left(\frac{\beta_{H}h^{\frac{1}{2}}(L)}{\epsilon}\right).
\end{equation*}
The purpose of this article is to show that this entropy is monotonous under a renormalization group flow; The RG-2 flow. This flow  makes to evolve the metric of the black hole, leading to an evolution of the black hole entanglement entropy. In two dimensions, when considering the non-linear sigma model with a Riemannian manifold as target space, the renormalization group flow has to be calculated perturbatively. Moreover, the target space of the non-linear sigma model can be a non-compact space, in particular, it can be a black hole. In the non-linear $\sigma$ model, the first loop aproximation of the RG flow is the Ricci flow. However, when curvature is high the Ricci flow is not a good approximation anymore \footnote{Due to the fact that a rigurous mathematical formulation of quantization is unavailable, it is not clear how big the contribution of moore loops would be.}. We should  consider the two loop beta functions, this new flow is known as the RG-2 flow \cite{Gimre:etal2}\cite{Branding:2015}. These intrinsic flows are natural candidates for studying the evolution of a black hole entanglement entropy. The evolution of this type of entropy under Ricci flow (1-loop) is studied in \cite{Solodukhin:2006ic} and here we extend the study to the two loop approximation, the RG-2 flow. \\
There are some results about the monotonicity of different entanglement entropies under a given RG flow. In \cite{Kim:2016jwu}, using an holographic technique the authors show how an entanglement entropy evolves along an RG flow. They found that in the UV limit, this entanglement entropy is not related  to thermal entropy\footnote{In the UV limit, entanglement entropy allows a thermodynamics-like law. Thus, introducing an entanglement temperature $T_{E}$ we have, ignoring all order corrections, the law: $\Delta S \approx \frac{\Delta E}{T_{E}}$. This similarity is only partial. Ignoring all higher order corrections, the entanglement temperature shows an universal behaviour inversely proportional to the subsystem size, in contrast to the thermodynamic temperature that is independent of the size, this why it cannot be reinterpreted as the thermodynamic law of a real thermal system.}. However, in the IR limit, the entanglement entropy can be decomposed into two parts. One is the contribution\footnote{This contribution is given by the thermalization of the excited state entanglement entropy.} that leads to the thermal entropy corresponding to the Bekenstein-Hawking entropy of the dual black hole geometry, the other is the remaining quantum entanglement near the entangling surface. Thus, we expect to find some kind of monotonicity under our RG-2 flow.
In order to study the evolution of the 2-dimensional Schwarzschild black hole entanglement entropy we have to be sure that the RG-2 flow have the desired properties, namely  if we have an initial asymptotically black hole metric it should remain asymptotically flat through the evolution. Moreover, the flow should not develop new horizons because of the evlotion. Finally, we have to show that the scalar curvature remains positive during the evolution. We will show that all these properties are satisfied by our flow and we will discuss about the extension of the results to higher loops, relating them to the weak parabolicity of the flow.\\
The Ricci flow has been used for proving the Thurston's geometrization conjecture\cite{Perelman:2006un,Perelman:2006up, Hamilton:1,Rflow:1}. For some applications for the Ricci flow to physics see \cite{Woolgar:2007vz,Carfora:2010iz,Headrick:2006ti, Husain:2008rg}. It is interesting to ask if the results obtained for the Ricci flow, are still satisfied when we make quantities evolve under a higher loop flow. Mathematically it is not easy to extend the results to higher order flows, but in two dimensions we have some results that will let us approach the problem safely.\\
In section \ref{bhentropy}  we make a brief review of 2-dimensional black holes and their entanglement entropy, in section \ref{theflow} we proceed to present the RG-2 flow set up and the black hole evolution. In section \ref{asymhaw} we present the results for the evolution of curvature, the existence of apparent horizons and ancient solutions. In section \ref{entropyev} we show that entanglement entropy is monotonically decreasing under RG-2 flow provided that the scalar curvature of the metric is positive at the horizon and given that we are inside the zone where the flow is weakly parabolic.
 
\section{2-dimensional  black holes and entanglement entropy}\label{bhentropy}

In this section we introduce the quantity that we are going to make evolve, namely the entanglement entropy in the background of a two dimensional euclidean Schwarzschild black hole. This metric corresponds to the analytical continuation of the two dimensional Schwarzschild metric. We start by giving a prescripton for calculating the entanglement entropy for a system that consists of two subsystems  $A$ and $B$ which are isolated form each other. Therefore, a quantum field $\psi$ will take values in both subsystems $\psi_{A}$ and $\psi_{B}$ respectively. If an observer have access only to one subsystem $A$ then we have to partially trace over the subsystem $B$ leading us to the definition of entanglement entropy for the zone $A$. We will adapt this recipe for calculating the entanglement entropy for the sub-zones created when a black hole horizon divides the spacetime background in two sub-zones, one inside the horizon and the other outside of it. Here we present a brief review of the calculation and the results that we are going to use in the  article, we  follow \cite{Solodukhin:2011gn}. See also \cite{Solodukhin:2006xv}.\\
Let us consider a two dimensional quantum field $\psi(x^{\mu})$ and chose cartesian coordinates $x^{\mu}=(t,x)$, where $t$ now represents the euclidean time. A surface $\Sigma$ defined by the condition $x=0$ and $t=0$ divides the hypersurface $t=0$ in two zones. Then, the vacuum state of $\psi$ satisfies in the boundary: $\psi(t=0,x)=\psi_{o}(x)$. Hence, the ground state is written as
\begin{equation}
\Psi(\psi_{o})=\int_{\psi(0,x)=\psi_{o}(x)} \mathcal{D}\psi e^{-W[\psi]},
\end{equation}  
where  $W[\psi]$ is the action of the field $\psi$. The two sections $x>0$ and $x<0$ separate naturally the boundary data, therefore we define:
\begin{eqnarray}
\psi_{-}(x)=\psi_{0}(x)\,\,\,\,with\,\, x<0,\\
\psi_{+}(x)=\psi_{0}(x)\,\,\,\,with\,\, x>0.
\end{eqnarray}
If we consider an observer outside of the $x<0$ zone we have to trace over the fields $\psi_{-}$, and therefore the density matrix can be written
\begin{equation}\label{densitym}
\rho(\psi_{+}^1,\psi_{+}^2)=\int \mathcal{D}\psi_{-}\Psi(\psi_{+}^1,\psi_{-})\Psi(\psi_{+}^2,\psi_{-}).
\end{equation}
The previous path integral is defined over the Euclidean spacetime except the cut $(t=0,x>0)$. It is clear that the field $\psi$ in the path integral takes the boundary values $\psi_{+}^1$ below the cut and $\psi_{+}^2$ above the cut. Thus, the trace of the density matrix (\ref{densitym}) is calculated by the euclidean path integral over the fields defined on a covering of the cut spacetime. Therefore, the $n-$th power of the density matrix is going to be a $n-$sheeted covering of the cut spacetime. From the geometrical point of view this $n-$fold is a flat cone $C_{n}$. Then, assuming that it is possible to continue analytically over any non-integer $n$ we can calculate the entanglement entropy by using:
\begin{equation}
S=(n\partial_{n}-1)W(n)\vert _{n=1}.
\end{equation}
We want to use the previous prescription for calculating the entanglement entropy in the background of the Euclidean Schwarzschild black hole in two dimensions. In a black hole the event horizon protects the interior from causal relation with the outside, therefore it divides the spacetime into two subsystems providing an ideal set up for applying the tools of entanglement entropy. Identifying all modes inside event horizon with the modes that are going to be traced over, and using the replica trick, entanglement entropy can be calculated.\\
We start by considering a two dimensional eternal\footnote{In an eternal black hole all definitions of horizon coincide.} black hole, a Schwarszchild black hole. This kind of black hole admits a maximally analytic extension. It also has a killing horizon: a null hypersurface on which the killing vector $\xi_{t}$ is null. In this black hole the future horizon intersects the past horizon in a bifurcation surface. It also admits an analytic continuation $T\rightarrow it$, where $T$ is the time coordinate of the spacetime and $t$ the euclidean time.\\
The ground state of our black hole is going to be defined by the euclidean path integral over fields defined on a half of the spacetime\footnote{This half instanton has a Cauchy surface on which coordinates can be chosen without problem. Moreover, we specify the boundary conditions, of our black hole wave function, in this Cauchy surface.} 
\begin{equation}\label{einstanton}
ds_{E}=\beta_{H}^2f(x)d\varphi^2+\frac{1}{f(x)}dx^2
\end{equation}
where $-\frac{\pi}{2}\leq\varphi \leq \frac{\pi}{2}$. Thus, for a black hole, the fields $\psi_{-}(x)$ and $\psi_{+}(x)$ are the boundary values defined on a Cauchy surface which is inside and outside respectively of the horizon $\Sigma$. 
For getting  a regular space there is a closing for the Euclidean time $t$, $0\leqslant t\leq \beta_{H}$, the period of the closing  is $\beta_{H}$. The quantity $\beta_{H}=T_{H}^{-1}$ is the inverse of the Hawking temperature $T_{H}$, and it can be determined by the derivative of the metric function $h(x)$ at the horizon:
\begin{equation}
\beta_{H}=\frac{4\pi}{h'(x_{h})}.
\end{equation}
Near $\Sigma$ the $n-$ cover of the euclidean black hole instanton (\ref{einstanton}) looks as a cone, where the tip  is at $\Sigma$ and with angle deficit $\gamma=2 \pi (1-\phi)$. Thus, the calculation of entanglement entropy is reduced to the calculation of a path integral over a gravitational background  with a conical singularity.\\
In 2 dimensions the entanglement entropy and its UV finite terms can be calculated explicitly. It can be easily done because of the existence of a conformal symmetry which helps for reproducing completely -for a conformal field theory (CFT)- the UV finite part of the corresponding gravitational effective action. When considering regular two-dimensional spacetimes the result is the non-local Polyakov action. For a two-dimensional CFT with central charge $c$, this action can be written in the form\footnote{The Ployakov action can be derived by integrating the conformal anomaly $\langle T^{(m)}\rangle=-\frac{c}{12}R$ from $\psi=0$ to $\psi$, using the fact that $\frac{\delta S_{eff}}{\delta\psi}=\frac{1}{2\pi}\sqrt{g}\langle T^{(m)}\rangle$.}
\begin{equation}\label{Polyakov}
Wp_{\mathcal{M}}=\frac{c}{48 \pi}\int_{\mathcal{M}}\left(\frac{1}{2}(\nabla \psi)^{2}+\psi R\right),
\end{equation}
where the field equation for $\psi$ is
\begin{equation}\label{psiequation}
\nabla^{2}\psi=R.
\end{equation}

As we already pointed out, we have to carry out the integral (\ref{Polyakov}) in a manifold $\mathcal{M}$ with a conical singularity with angle deficit $\gamma$, whence, the Polyakov action becomes:
\begin{equation}\label{Polyakov:1}
Wps_{\mathcal{M}}=\frac{c}{48 \pi}\int_{\mathcal{M}}\left(\frac{1}{2}(\nabla \psi)^{2}+\psi R\right)+\frac{c}{12}(1-\phi)\psi_{h}+o(|1-\phi|^2),
\end{equation}
where $\psi_{h}$ represents the value that $\psi$ takes at the horizon. Applying the replica trick to the action (\ref{Polyakov:1}) the entanglement entropy contribution from the UV finite part of the effective action is obtained. In a conformal gauge $g_{\mu\nu}=e^{2\sigma}\delta_{\mu\nu}$ where $\psi=2\sigma$ the complete entanglement entropy is 
\begin{equation}\label{ententropy}
S=\frac{c}{6}\sigma_{h}+\frac{c}{6}\ln \frac{\Lambda}{\epsilon},
\end{equation}
where $\Lambda$ is the infrared cut-of, and $\sigma_{h}$ represents $\sigma$ evaluated at the horizon \cite{Solodukhin:2006xv}.\\
We are going to use the formula (\ref{ententropy}) and calculate the entanglement entropy of a  2-dimensional Euclidean black hole geometry described by the metric
\begin{equation}\label{metric1}
ds^{2}=h(x)dt^{2}+\frac{1}{h(x)}dx^{2},
\end{equation}
where $h(x)$ has a zero in $x=x_{h}$. We also put the black hole in a box of size $L$, then, $x_{h}\leq x\leq L$. With the use of the change of variable  
\begin{equation}
\ln z=\frac{2\pi}{\beta_{H}}\int_{L}^{x}\frac{dx}{h(x)},
\end{equation}
we can see that the metric (\ref{metric1}) is conformal to the flat disk of radius $z_{o}$, indeed the metric becomes
\begin{equation}
ds^{2}=e^{2\sigma}z_{o}^{2}(dz^{2}+z^{2}d\bar{t}^{2}),
\end{equation}
where
$\sigma=\frac{1}{2}\ln h(x)-\frac{2\pi}{\beta_{H}}\int_{L}^{x}\frac{dx}{h(x)}+\ln\frac{\beta_{H}}{2\pi z_{o}}$ and $\bar{t}=\frac{2\pi t}{\beta_{H}}$. Therefore, the calculation of entanglement entropy can be done analytically using (\ref{ententropy}), we only need the factor $\sigma$ evaluated at the horizon $x_{h}$. After some arrangements the entanglement entropy of the  2-dimensional black hole (\ref{metric1}) becomes \cite{Solodukhin:2006xv}:
\begin{equation}\label{ententropy:1}
S=\frac{c}{12}\int_{x_{h}}^{L}\frac{dx}{h(x)}\left(\frac{4\pi}{\beta_{H}}-h'\right)+\frac{c}{6}\ln\left(\frac{\beta_{H}h^{\frac{1}{2}}(L)}{\epsilon}\right),
\end{equation}
where the term depending on ($\Lambda$, $z_{o}$) has been omitted and $\epsilon$ is the UV regulator\footnote{Defining the  coordinate invariant size (of the subsystem outside the black hole) as $L_{I}=\int_{x_{h}}^{L}\frac{dx}{\sqrt{h(x)}}$ it was shown that the  entanglement entropy (\ref{ententropy:1}) is an increasingly function of $L_{I}$.}.  This is the entropy that we want to make evolve under the RG-2 flow, we will see that we have to make some arrangements before applying the evolution equation, and we also have to be sure, all the time, if we are in the zone where the flow is weakly parabolic.

\section{The RG-2 flow}\label{theflow}
Here we introduce the RG-2 flow in two dimensions. Although the mathematical literature refers mostly to the compact case  we will adapt the results to asymptotically flat manifolds. 
When perturbately quantizing the action of non-linear sigma model, it is necessary to introduce a momentum cut-off $\Lambda$. After completing the whole procedure we obtain the renormalization group flow equations. Then, setting  $\lambda:=-\ln (\frac{\Lambda}{\Lambda'})$ the renormalization group flow equations for the non-linear sigma model in two dimensions can be written: 

\begin{equation}\label{renorm:1}
\frac{\partial g_{ij}}{\partial \lambda}=-\beta_{ij}(g).
\end{equation}

We will have different $\beta_{ij}$ functions depending of the number of loops considered. Since in two dimensions the Riemann tensor can be written as $R_{i j kl}=g_{i[k}g_{l]j}R$ we have, after a re-escalation, for the two loop approximation:

\begin{equation}\label{RG2floweq}
\partial_{\lambda}g_{ij}=-Rg_{ij}-\frac{\alpha}{4}R^{2}g_{ij},
\end{equation}
here $R$ is the Ricci scalar of the metric $g_{ij}$.
The flow defined by (\ref{RG2floweq}) is called the \textbf{RG-2 flow} and it is the two loop approximation to the RG flow of the non-linear sigma model. The flow in (\ref{RG2floweq}) is not parabolic, the term $\frac{\alpha}{2}R^{2}g_{ij}$ spoils the parabolicity of the flow. Nevertheless, it has been shown that there are regions where the flow is weakly parabolic, namely the regions where $1+\frac{\alpha}{2}R>0$ ~\cite{Oliynyk:2009rh}. Moreover, the existence of zones of weak parabolicity for the RG-2 flow can be extended to higher dimensions \cite{Gimre:2014jka,Gimre:etal2}.  These regions are clearly determined, and we must work inside them to ensure the existence of solutions.

\subsection{Black hole evolution}

We want to make evolve a black hole metric in two dimensions. In order to do that, we take the same ansatz as given in \cite{Solodukhin:2006ic} and  at "time" ($\lambda=0$) we chose a metric of the Schwarzschild type:
\begin{equation}\label{initialm}
ds^{2}=f(x)dt^{2}+\frac{1}{f(x)}dx^{2},
\end{equation}  

where $f(x) $ is positive everywhere but becomes zero when evaluated at the horizon. When the black hole is non-degenerated near the horizon, $x=x_{h+}$, we can write
\begin{equation}
f(x)=q_{o}(x-x_{h+})+O((x-x_{h+})^{2}),
\end{equation}
thus, the Hawking temperature  $T_{H}=\frac{q_{o}}{4\pi}$ is different from zero. Moreover, $f\rightarrow 1 $ when $x\rightarrow \infty$, which implies that the metric is asymptotically flat.

Now we define the  metric for $\lambda > 0$ :
\begin{equation}\label{metric:2}
ds^{2}=f(x,\lambda)dt^{2}+\frac{e^{2\phi(x,\lambda)}}{f(x,\lambda)}dx^{2},
\end{equation}
then, its Ricci scalar is 
\begin{equation}
R=e^{-2\phi}(-\partial_{xx}f+\partial_{x}\phi\partial_{x}f),
\end{equation}
and we set the initial value  $\phi(x,0)=0$.\\

Replacing (\ref{metric:2}) in the RG-2 flow  (\ref{RG2floweq}) we obtain:

\begin{eqnarray}
\frac{\partial f}{\partial \lambda}=-e^{-2\phi}(-\partial_{xx}f+\partial_{x}\phi\partial_{x}f)f-\frac{\alpha}{4}e^{-4\phi}(-\partial_{xx}f+\partial_{x}\phi\partial_{x}f)^2 f\label{RG:1}\\
\frac{\partial e^{2\phi}}{\partial \lambda}=-2(-\partial_{xx}f+\partial_{x}\phi\partial_{x}f)-\frac{\alpha}{2} e^{-2\phi}(-\partial_{xx}f+\partial_{x}\phi\partial_{x}f)^2\label{RG:2}
\end{eqnarray}
The previous system of equations, except for the $\alpha$ terms, is the same one obtained in the Ricci flow case. However, now the type of equations is completely different, they are fully non-linear and weakly parabolic in the zone $1+\frac{\alpha}{2}R>0$.
From (\ref{RG:1}) and (\ref{RG:2}) we get
 \begin{equation}\label{equationf}
\frac{\partial f}{\partial \lambda}=\frac{\partial \phi}{\partial \lambda}f,
 \end{equation}
 which is the same equation obtained for the Ricci flow (see eq. (3.6) on ~\cite{Solodukhin:2006ic}) then, using the initial condition (\ref{initialm}) we get from the previous equation
 \begin{equation}\label{phi}
 \phi=\ln\left(\frac{f(x,\lambda)}{f(x,0)}\right).
 \end{equation}
Using the eq. (\ref{phi}) we can write the metric (\ref{metric:2}) as:

\begin{equation}\label{metric:3}
ds^{2}=f(x,\lambda)\left(dt^{2}+\frac{dx^2}{f^{2}(x,0)}\right),
\end{equation}
Our first result is that both flows admit the same initial metric, the same result will also holds for higher loops.\\
The function $f(x,\lambda)$ satisfies the equation
\begin{equation}
\frac{\partial f}{\partial \lambda}=f(x,0)\partial_{x}\left(\frac{f(x,0)\partial_{x}f}{f}\right)-\frac{\alpha}{4}\frac{f^{2}(x,0)}{f}\left(\partial_{x}\left(\frac{f(x,0)\partial_{x}f}{f}\right)\right)^{2}.
\end{equation}
Introducing a new coordinate $r$  defined by~\cite{Solodukhin:2006ic}
\begin{eqnarray}\label{changevar}
r=\int^{x}_{x_{h}}\frac{f(x,\lambda)}{f(x,0)}dx\label{coord:1}\,,\,\,\,\,\,\,\,\,\,\,\,\,\,\,\,\,\,
\frac{\partial r}{\partial\lambda}=\partial_{r}f-q_{o},\label{coord:2}
\end{eqnarray}
where $r=0$ corresponds to $x=x_{h}$, the metric (\ref{metric:3}) becomes:
\begin{equation}\label{metric:1}
ds^{2}=f(r,\lambda)dt^{2}+\frac{1}{f(r,\lambda)}dr^{2}.
\end{equation}

Now we are ready to make evolve the metric (\ref{metric:1}) under the RG-2 flow  (\ref{RG2floweq}):

\begin{equation}\label{RG2flow:1}
\frac{\partial f}{\partial \lambda}=f \partial_{rr}f-\frac{\alpha}{4}f(\partial_{rr}f)^{2}-(\partial_{r}f)^{2}+q_{o} \partial_{r}f,
\end{equation}

here $q_{o}=\partial_{\lambda}f(x_{h+},\lambda)\vert_{\lambda=0}$. 

The previous equation can be written as:

\begin{equation}\label{RG2flow}
\frac{\partial g_{ij}}{\partial\lambda}=-Rg_{ij}-\frac{\alpha}{4}R^{2}g_{ij}-\nabla_{i}\eta_{j}-\nabla_{j}\eta_{i},
\end{equation}
with 
\begin{equation}\label{vector:1}
\eta_{t}=0,\,\,\,\,\eta_{r}=\frac{1}{f(r,\lambda)}(\partial_{r}f-q_{o}).
\end{equation}
Setting\footnote{For metric (\ref{metric:3}) we have
\begin{equation}
\psi(x,\lambda)=\ln \left(\frac{f(x,0)}{f(x,\lambda)}\right)+\psi_{o}(x).
\end{equation}}
\begin{equation}
\psi(r,\lambda)=\int^{r}\frac{d\bar{r}}{f(\bar{r})\lambda}(q_{o}-\partial_{r}f),
\end{equation}
we can write
\begin{equation}
\eta_{t}=\partial_{t}\psi(r,\lambda),\,\,\,\,\,\eta_{r}=-\partial_{r}\psi(r,\lambda).
\end{equation}
This $\psi$ is a solution to the equation\footnote{The entanglement entropy can be written, with an appropriate choice of the integration constants, as:
\begin{equation}
S=-\frac{c}{12}\psi(0)+\frac{c}{6}\ln(\frac{1}{\epsilon}).
\end{equation}} (compare with equation (\ref{psiequation})):

\begin{equation}
\Delta \psi=-\partial_{rr}f.
\end{equation}

In this section we have extended the analysis made in \cite{Solodukhin:2006ic} to the RG-2 flow case. In this particular coordinates it is easy to see the general form of RG-2-DeTurck flow, this flow is parabolic (making the RG-2 flow weakly parabolic). We see that the extension is straightforward, and therefore, it can be done for higher loops. The only problem will be the existence of solutions for these flows, which has not been determined yet.  

Now, let's see how some properties of our two dimensional black hole behave under RG-2 flow evolution.

\section{Asymptotical behaviour and Hawking temperature}\label{asymhaw}
An asymptotically flat metric will remain asymptotically flat after evolution under the RG-2 flow. Indeed, an asymptotically flat metric $g_{ij}$ at infinity behaves as $g_{ij}=1+\eta_{ij}$ with $\eta_{ij}<<1$, then $\eta_{ij}$ evolves as
\begin{equation}
\frac{\partial \eta}{\partial \lambda}=\partial_{rr}\eta+q_{o}\partial_{r}\eta-\frac{\alpha}{4}\eta(\partial_{rr}\eta)^{2}-(\partial_{r}\eta)^2.
\end{equation}

If we assume that $\eta_{o}(r)=\frac{1}{r^{k}}, k\geq 1$ when $r$ is large,  we can neglect the terms $(\partial_{rr}\eta)$ and $(\partial_{rr}\eta)^{2}$, then $\eta(r,\lambda)$ satisfies:
\begin{equation}
\frac{\partial \eta}{\partial \lambda}=q_{o}\partial_{r}\eta.
\end{equation}
Using the initial value $\eta_{o}(r)$ it is straightforward to obtain $\eta(r,\lambda)=\eta_{o}(r+q_{o}\lambda)\approx \frac{1}{(r+q_{o}\lambda)^{k}}$. This shows that $\eta(r,\lambda)$ will decay for large $r$ and large $\lambda>0$. The bound obtained for the Ricci flow $|\eta(r,\lambda)|\leq |\eta_{o}(r)|$ is still satisfied when evolution is under the RG-2 flow. On the other hand, we know that $R=-\partial_{rr}\eta(r,\lambda)$ then  the asymptotic flatness is preserved by the RG-2 flow. This result was expected since the terms corresponding to the second loop correction at infinity are small compared with those of the one loop flow. Adding more loops will not change this behaviour, and therefore we can state that a two dimensional asymptotically  flat  black hole will remain asymptotically flat for all loops.\\
Now we analyse what happens when we are near the horizon. We assume that near $r=0$ the function $f(r,\lambda)$ can be expanded using a Taylor series, 
\begin{equation}\label{series}
f(r,\lambda)=\sum_{i=0}^{\infty}a_{i}(\lambda)r^{i}.
\end{equation}
We replace the series (\ref{series}) in (\ref{RG2flow:1}) and we will obtain the flow equations for the coefficients of the series. In general, we will obtain the Ricci flow equations plus the $\alpha-$ terms corresponding to the correction introduced by the RG-2 flow.  The flow equations for the first coefficients become:
\begin{eqnarray}
\frac{\partial a_{o}}{\partial\lambda}&=&2a_{o}a_{2}+a_{1}(q_{o}-a_{1})-\alpha a_{o}a_{2}^{2}\\ \frac{\partial a_{1}}{\partial\lambda}&=&6 a_{o}a_{3}+2a_{2}(q_{o}-a_{1})-\alpha(a_{1}a_{2}^2+3a_{o}a_{3})
\end{eqnarray}

In general (when $n \geq 2$)
\begin{equation}
\begin{split} 
\frac{\partial a_{n}}{\partial\lambda}&=(n+2)(n+1)a_{o}a_{n+2}+(n+1)(q_{o}-a_{1})a_{n+1}+(n^2-1)a_{1}a_{n+1}\\
&+\sum_{m=2}^n \Bigg(m(2m-n-3)a_{m}a_{n-2+2}\\
&-\frac{\alpha}{4}\sum_{k=0}^{m}(m+2)(m+1)(m+2-k)(m+1-k)a_{m+2}a_{m-k+2}a_{n-m}\Bigg).
\end{split}
\end{equation}

We know that when $\lambda=0$ we have $a_{o}=0$ and $a_{1}=q_{o}$. Therefore,
\begin{equation}
\frac{\partial^{k} a_{o}}{\partial \lambda^{k}}=0.
\end{equation}
The previous equation implies that $a_{o}=0$ for all $\lambda>0$. The horizon will be located always at $r=0$. But contrary to what happens with the Ricci flow now we have
\begin{equation}
\frac{\partial a_{1}}{\partial \lambda}=-\alpha q_{o}a_{2}.
\end{equation}
Thus, we can see that $a_{o}=0$  for all $\lambda$, and $a_{1}=q_{1}$ at $\lambda=0$. The $a_{1}$ coefficient evolution will depend on the evolution of  the other coefficients. This will also happen with Ricci flow evolution, but for higher $n$. This is a manifestation of the higher curvature terms. Now, we know that $T_{H}=\frac{1}{4\pi}\partial_{r}f(0,0)$ is the Hawking temperature of the black hole metric \eqref{metric:3}. Therefore, the Hawking temperature remains unchanged with evolution, this result holds for any loop\footnote{The Hawking temperature, for the kind of euclidean black hole that we are considering, only depends on the asymptotic limit of the metric, which as we have shown does not change. Indeed, from equation \eqref{metric:3} we can see that the factor $f(x,\lambda)$ is the only one in the metric depending on $\lambda$  therefore, the temperature is fixed for every lambda.}.

\section{Ancient solutions, scalar curvature, and horizons}\label{scalarcurv}

The appearance of singularities is a big problem when dealing with intrinsic flows, such as the RG-2 flow. It is not easy to control the behaviour of curvature when the manifold is evolving. Here we will study how the scalar curvature  behaves under the RG-2 flow\footnote{We also want to study the Kretschmann scalar in order to see if there is a curvature singularity. However, in two dimensions the Kretschmann scalar is $R_{ijkl}R^{ijkl}=R^2$, then controlling the evolution of $R$ is enough to ensure the evolution without singularities.}.
\subsection{Evolution of scalar curvature}
The scalar $R$ evolves under the  eq. (\ref{RG2flow}) as
\begin{equation}\label{scalarevolve}
\frac{\partial R}{\partial \lambda}=(1+\frac{\alpha}{2}R) \Delta{R}+\frac{\alpha}{2}|\nabla R|^{2}+\left(1+\frac{\alpha}{4} R\right)R^{2}+\eta^{r}\partial_{r}R.
\end{equation}
Using the metric eq. (\ref{metric:1}) together with (\ref{vector:1}) the previous equation transforms to:
\begin{equation}\label{curv}
\frac{\partial R}{\partial \lambda}=(1+\frac{\alpha}{2}R)f \partial_{rr}R+\frac{\alpha}{2}|\partial_{r}R|^{2}+\left(1+\frac{\alpha}{4}R\right)R^{2}+q_{o}\partial_{r}R.
\end{equation}

We have to be sure that the flow does not change the sign of the curvature during the evolution. If we start with a manifold of positive scalar curvature the flow should maintain its positivity. This requirement is needed in order to stablish monotonicity.\
If we assume that $R(r,\lambda)$ has a local minimum at some $r_o$, then $\partial_r R(r_{o},\lambda)=0$ and $\partial_{rr}R(r_{o}\lambda)>0$. If the the scalar curvature $R$ is positive then $1+\frac{\alpha}{2}R>0$, which is the condition needed in order to have a weakly parabolic flow. Moreover, the positiveness of the scalar curvature implies that $(1+\frac{\alpha}{4}R)>0$, and therefore we can conclude that $\partial_{\lambda}R(r_{o}\lambda)>0$. It implies that the point $r_{o}$ is a global minimum, the scalar curvature will increase from $r_o$ and therefore, it will remain positive during the evolution.  This result was expected and it was already know for two dimensional closed manifolds. However, in our case we have an asymptotically flat manifold with a horizon, therefore we have to look what is happening at these important points\footnote{ The RG-2 flow in two dimensions is weakly parabolic only when $1+\frac{\alpha}{2}R>0$, then a maximum principle can be used in equation (\ref{curv}). Usually, a zone of parabolicity and backwards parabolicity  are defined $\mathcal{M}_{-}=\{g\in \mathcal{M}\mid 1+\frac{\alpha}{2}R>0\}$  and $\mathcal{M}_{+}=\{g\in \mathcal{M}\mid 1+\frac{\alpha}{2}R<0\}$ respectively, where $\mathcal{M}$ is the space of smooth Riemannian metrics. This analysis shows that, even in the zones where the curvature scalar is negative it will remain negative \cite{Oliynyk:2009rh,Branding:2015}.}.\\
At infinity the manifold is flat, then $R=0$ when $r\rightarrow \infty$, so it remains no negative, and as we have seen in section \ref{asymhaw} it will remain asymptotically flat. On the other side, in the horizon, the function $f$ vanishes and therefore we have that

\begin{equation}\label{scalarev0:1}
\frac{\partial R(0,\lambda)}{\partial \lambda}=\frac{\alpha}{2}|R^{'}(0,\lambda)|^{2}+\left(1+\frac{\alpha}{4}R(0,\lambda)\right)R^{2}(0,\lambda)+q_{o}R'(0,\lambda)
\end{equation}

First we assume that $\partial_{r}R(0,\lambda)>0$ then, in the zones where $1+\frac{\alpha}{4}R(0,\lambda)>0$, because of (\ref{scalarev0:1}), we have $\partial_{\lambda}R(0,\lambda)>0$, therefore $R(0,\lambda)$ is increasing function of $\lambda$, then $R(0,\lambda)$ remains positive during evolution. Now we assume the opposite, namely $\partial_{r}R(0,\lambda)<0$. Then, we have two options: if there is a local minimum at some point $r_{o}>0$ then $R(0,\lambda)>R(r_{o},\lambda)>0$. If there is not such a minimum. then $R(0,\lambda)>R(\infty,\lambda)=0$. Thus, $R$ remains positive everywhere including the point $r=0$. For compact surfaces a similar result has been found for the $RG_{o}$-2 flow\footnote{The $RG_{o}$-2 flow is the RG-2 flow eq. (\ref{RG2flow}) but without the DeTurck term.}.  

\subsection{Ancient solutions}

The stationary points of the flow are easily characterized. From (\ref{RG2floweq}) we see that the fixed points of the flow are given by the spaces of constant curvature. The spaces such that $R=0$ and $R=-\frac{4}{\alpha}$ are the IR and UV fixed points respectively. The ancient solutions are trickier. If we define 
\begin{equation}\label{solfix}
f_{as}(\lambda,x)=\frac{f_{R}(x)}{h(\lambda)},
\end{equation}
where $f_{R}(x)$ is a metric of constant negative curvature ($R=-1$). The scalar curvature of $f_{as}(\lambda,x)$ is $R(\lambda)=h(\lambda)$. Then  from (\ref{scalarevolve}) we obtain
\begin{equation}
\frac{dh}{d\lambda}=h^{2}+\frac{\alpha}{4}h^3
\end{equation}
whose general solution is
\begin{equation}
\lambda(h)=-\frac{1}{h}-\frac{\alpha}{4}\ln(4h+\alpha h^{2})+c
\end{equation}
The solution (\ref{solfix}) was found in \cite{Oliynyk:2009rh}, and it corresponds to an eternal solution connecting two fixed points, the IR fixed point and the UV fixed point. It is interesting to note that the solution $f_{as}$ for negative $\lambda$ belongs to the zone where the flow is backwards parabolic, and for positive $\lambda$ it belongs to the zone where the flow is parabolic.

\subsection{Horizons}
We are interested in studying the appearance of horizons. In section \ref{bhentropy} we used the fact that in two dimensions, because of the conformal symmetry, the only quantity that we need in order to calculate the entanglement entropy in an Euclidean black hole is the conformal factor $\sigma$ of the metric evaluated at the horizon. In order our setup to be consistent we must ensure that our metric does not develop new horizons\footnote{Evidently, if the black hole metric develop new horizons the calculation of entanglement entropy presented in section \ref{bhentropy} does not work anymore.}. Moreover, the topology of an Euclidean black hole is different from a Lorentzian one. In the Euclidean case, the time is periodic $0\leqslant t\leq \beta_{H}$, and therefore any observable that we calculate using the Euclidean path integral will be periodic in imaginary time. This periodicity helps to avoid a conical singularity at the horizon, leading to have tip at the horizon. Therefore, the geometry of the Euclidean Schwarzschild black hole has the shape of a "cigar" with a tip at the horizon. We will maintain the name horizon, inherited form their Lorentzian counterpart, although the physics now is different. Notwithstanding the conceptual change both points remain connected by Wick rotation, therefore when we have a horizon in our Euclidean manifold by reverse subtitution we can find that this point correspond to a horizon in the Lorentzian manifold. Thus, if our Euclidean metric does not develop a new horizon because of evolution we can conclude that the corresponding Lorentzian metric neither does it. Thanks to the change of variable (\ref{changevar}) the horizon in our Euclidean Shchwarzschild black hole is going to be located at $r=0$. \\
As we showed in the previous sections  the flow maintains asymptotically flatness and positiveness of the scalar curvature. An asymptotically flat initial metric $f(r,0)$ will maintain this property, and therefore at infinity, after evolution, it will be $f(\infty,\lambda)=1$. If it has a horizon, let's say at $r=0$ then $f(0,\lambda)=0$. Now if we assume that for  some $\lambda>0$ it appears a point $r=r_{1}>0$ such that $f(r_{1},\lambda)=0$ (a new horizon), then it will be an extreme point $f$ and therefore, $\partial_{r}f(r_{1},\lambda)=0$. We have two cases: if it is a minimum $\partial_{rr}f(r_{1},\lambda)>0$. The eq. (\ref{RG2flow:1}) gives 
\begin{equation}\label{restriction:1}
\frac{\partial f}{\partial \lambda}=f\partial_{rr}f\left(1-\frac{\alpha}{4}\partial_{rr}f\right).
\end{equation}
From the previous equation we can see that if $1-\frac{\alpha}{4}\partial_{rr}f>0$ we can conclude that $\partial_{\lambda}f>0$, which implies that $f(r_{1},\lambda)$ is increasing with $\lambda$, therefore $f$ will never develop a horizon. We have assumed that $f$ belongs to the zone where $1-\frac{\alpha}{4}\partial_{rr}f>0$, so let us take a look more closely to this condition.\\
We know that $R=-\partial_{rr}f$, therefore the zone we are analysing becomes
\begin{equation}\label{zone2}
1+\frac{\alpha}{4}R>0.
\end{equation} 
Inequality (\ref{zone2}) will be satisfied wherever $R$ is positive. Following our argument we have that $\partial _{rr}f>0$, then the scalar curvature has to be negative. If it is negative it can only be in the zone where $-\frac{4}{\alpha}<R<0$, where we know that the flow is parabolic and the curvature is controlled.\\
The second case is when $r_{1}$ is a maximum, then $\partial_{rr}f(r_{1},\lambda)<0$. Inequality (\ref{zone2}) will be satisfied because the scalar curvature $R$ is going to be positive. Thus, from equation (\ref{restriction:1}) we get that $\partial_{\lambda}f<0$, it implies that the metric is a decreasing function of $\lambda$. A similar result in the context of General Relativity was found in \cite{LassoAndino:2018zhb}.

\section{Entropy}\label{entropyev}
Here we present the calculation of entropy evolution under RG-flow. From the mathematical point of view there have been some definitions for entropy. The most famous one is the surface entropy $N(g)=\int_{M}R \log R$. In \cite{Rflow:1} it is shown  that the surface entropy, sometimes called Hamilton's entropy, is monotonous under the Ricci flow when  $M$ is a compact surface. The result was extended to asymptotically flat spaces in \cite{Solodukhin:2006ic}. The motivation behind this result is that our metric (\ref{metric:1}) is also a solution of the field equations derived from $N(g)$.\\
Similarly, we can think about extending this result to the RG-2 flow even to the first $\alpha$-correction. In \cite{Branding:2015} there is a proposal for a surface entropy:
\begin{equation}\label{surfentropy}
N_{\alpha}(g)=\int \left(R\log R+\frac{\alpha}{2}R^{2}\right)d\mu.
\end{equation}
Assuming $R>0$ it is shown that the second order surface entropy of a closed manifold is monotonous under the normalized RG-2 flow. The result is directly applicable to the RG-2 without the normalization. Indeed, defining $X_{i}=\nabla_{i}R+R\nabla_{i}f+\frac{\alpha} R\nabla_{i}R$ and calling $M_{ij}$ to the trace-free part of the Hessian of $\psi$, it can be shown that
\begin{equation}\label{surfentropyflow}
\frac{dN_{\alpha}}{d\lambda}=-\int_{M}\frac{|\nabla R+ R\nabla V+\frac{\alpha}{2}R\nabla R|^2}{R}-2\int_{M}|M_{ij}|^2<0.
\end{equation}
This result can be easily extended to asymptotically flat spaces. We consider a metric of the type (\ref{metric:1}) and $f(r)=1+O(r^{-k})$, with $k>0$ and $r$ big. The scalar curvature behaves as $R\sim 1/r^{k+2}$, then $R^{2}\sim 1/r^{(2k+4)}$. Moreover $R'\sim 1/r^{k+3}$, therefore the integrand in (\ref{surfentropy}) goes as $O(\ln r/r^{k+2})$. It is straightforward to calculate the convergence of the integrals in (\ref{surfentropyflow}). Thus
\begin{eqnarray}
\frac{|\nabla R+ R\nabla V+\frac{\alpha}{2}R\nabla R|^2}{R}\sim \frac{1}{r^{k+4}},\,\,\,\,\,\,\,\,\,\,\,\,\,|M_{ij}|\sim\frac{1}{y^{2(k+1)}}.
\end{eqnarray}
The integrals (\ref{surfentropy}) and (\ref{surfentropyflow}) converge. With all said we can assure that the Hamiltons's entropy is strictly decreasing (except when $f$ represents a steady soliton) under RG-2 flow  for asymptotically spaces.

\subsection{Entanglement entropy}
The black hole entanglement entropy is an of-shell definition, therefore it can be calculated for any black hole metric (not necessary a solution of equations of motion coming from any action). The entanglement entropy (\ref{ententropy:1}) in the new coordinate given by (\ref{changevar}) becomes
\begin{equation}\label{entropy:1}
S(\lambda)=\frac{c}{12}\left(\int_{0}^{L_{r}}dr\frac{(q_{o}-\partial_{r}f(r,\lambda))}{f(r,\lambda)}+\ln f(L_{r},\lambda)-2 \ln(q_{o}\epsilon)\right),
\end{equation}

where $\partial{r}f(0,\lambda)=q_{o}$ and as before $L_{r}$ is the value of the $r$ coordinate  at the boundary of the box. Then, the variation of the entropy under $\lambda$  is given by \cite{Solodukhin:2006ic}

\begin{equation}
\frac{\partial S(\lambda)}{\partial\lambda}=\frac{c}{12f(\delta)}\left(\frac{\partial f(\delta,\lambda)}{\partial\lambda}+q_{o}(\partial_{r}f(\delta,\lambda)-q_{o})\right).
\end{equation}
The proper variation of the entropy will be recovered in the limit $\delta\rightarrow 0$. The previous equation can be used for any intrinsic flow, in \cite{Solodukhin:2006ic} the evolution under the Ricci flow was studied. Here we extend the analysis to the RG-2 flow, and in principle,  the analysis can be extended to any loop approximation of the renormalization group flow.

Using (\ref{RG2flow:1}) at $r=\delta$ we find the variation of the black hole entanglement entropy under the RG-2 flow:
\begin{equation}
\frac{\partial S(\lambda)}{\partial\lambda}=\frac{c}{12}\left(\partial_{rr}f(\delta)-\frac{(\partial_{r}f(\delta)-q_{o})^{2}}{f(\delta)}-\frac{\alpha}{4}(\partial_{rr}f)^2+q_{o}\frac{\partial_{r}f}{f}\right)
\end{equation}
When we take the limit $\delta\rightarrow 0$ the terms that are of the order of $\delta$ vanish, and  we get
\begin{equation}
\frac{\partial S(\lambda)}{\partial \lambda}=\frac{c}{12}\left(\partial_{rr}f(\delta)-\frac{\alpha}{4}(\partial_{rr}f(\delta))^2\right)
\end{equation}
Since $R(0,\lambda)=-(\partial_{rr}f)(0,\lambda)$ we obtain

\begin{equation}\label{Rg2entropy}
\frac{\partial S(\lambda)}{\partial \lambda}=-\frac{c}{12}\left(R(0,\lambda)+\frac{\alpha}{4}R(0,\lambda)^2\right)
\end{equation}

The previous equation shows that if the scalar curvature is positive at the horizon for all $\lambda$, the entanglement entropy is monotonically decreasing under the RG-2 flow.   This is one of the main results of this work.  If we start with a metric with positive scalar curvature, it will maintain the positivity through the evolution as we have seen in section \ref{scalarcurv}. On that section, the requirement is the positivity of the term $\left(1+\frac{\alpha}{4} R\right)$, this is the only condition needed, and it appears again in (\ref{Rg2entropy}). For higher loops the pattern will be recognizable, however more conditions will arise because of the weak parabolicity of the flow, putting restrictions to the zones where the flow has solutions\footnote{The flow needs to maintain the positivity only at the horizon.}.

\section{Final comments}
The irreversibility of a process carries with itself the definition of entropy. When considering a general bosonic non-linear sigma model we find that the RG flow is an evolution equation for the metric, and at first order it coincides with the Ricci flow. A first thing to ask  about is if the same results hold for higher order loops. Here we have studied some properties of the RG-2 flow in two dimensions for the Euclidean Schwarzschild black holes. Some of them where expected, and are satisfied almost trivially, but others clearly reflect the influence of the $\alpha$-correction. For example, an asymptotically flat metric at the beginning of the flow remains asymptotically flat for all times, and since higher loops will add only small neglecting terms this result can be stated for all loops. In the compact case the end point of the flow, if any, are metrics of constant curvature, which are stationary points of the flow.  The Hawking temperature of the two dimensional euclidean static black hole does not change under the RG-2 flow. This result can be extended to higher loops. The Hawking temperature depend only on the asymptotic limit of the metric, the only requirement is the flow being smooth.\\
There are non trivial results, the existence of horizons is restricted to a given zone. However, if we consider a manifold with scalar curvature $R>0$ the restriction is evidently satisfied, it suggests that for higher loops the study of the conditions for the existence of horizons will lead us to find a condition for the existence of solutions to the flow.\\
The flow could lead to the appearance of a new horizon if we start evolving the flow with a metric outside of the zone $1+\frac{\alpha}{4}R>0$. Even inside that zone, when the scalar curvature $R$ is negative, horizons could appear. When studying higher flows it will appear more restrictions, and all of them will be related with the zone where the flow is parabolic, and the zones will be restricted more and more. However, it seems that the pattern will be recognizable.
The entanglement entropy of a two dimensional Schwarzchild black hole is monotonous under the RG-2 flow, the conditions needed to be satisfied are the positiveness of the scalar curvature at the horizon, and the scalar curvature satisfying $1+\frac{\alpha}{4}R>0$  at the horizon.\\
There is also another important aspect of the evolution. Almost all theorems regarding the Ricci flow (and the RG-2 flow) are stated for Riemannian manifolds, and certainly our black hole is Euclidean, therefore we can apply all results regarding Riemannian manifolds. However, the physics is different in both cases. In the Euclidean case the horizon is located at the tip of the "cigar" metric\footnote{In string theory, this metric is known as the Witten black hole.}, and thanks to the change of variable (\ref{changevar}) we have that the horizon in our Euclidean metric is located at $r=0$. We can recover the Lorentzian metric by reversing the Wick rotation, and the point $r=0$ will correspond to a horizon in the Lorentzian sense, that is why we have maintained the word horizon for the tip of the cigar.  This led us to say that if a horizon does not appear in the Euclidean metric it will not arise in the Lorentzian metric.\\
Finally, there are certain mathematical results that allow to extend what we have found for Riemannian manifolds to a higher dimensional Lorentzian manifolds. In other words, given a Riemannian metric we can built a Lorentz manifold by using this Riemannian metric, and the Ricci flow evolution of the new Lorentzian metric is well defined. See Theorem 2.1 in \cite{Chow:1995}. The new Lorentzian metric will have one dimension (corresponding to time) more. Therefore, the singularities appearing in the Riemannian metric can be realized in a particular Lorentzian manifold. Moreover, the existence of singularities in the Riemannian metric could lead to the existence of horizons in the Lorentzian metric. This aspect, in the of evolution of spacetime metrics under intrinsic flows, has not been studied extensively \cite{Chow:2006} and could be a problem to address in the future.

%%%%%%%%%%%%%%%%%%%%%%%%%%%%%%%%%%%%%%%%%%%%%%%%%%%%%%%%%%%%%%%%%%%%%%%%%%%%%%%%%%%%%%%%%%%%%%%%%%%%%%%%%%%%%%%%%%%%%%%%%%%%%%%%%%%%%%%%%%%%%%%%%%%%%%%%%%%%%%%%%%%%%%%%%%%%%%%%%%%%%%%%%%%%%%%

\section*{Acknowledgments}

This work was has been partially supported by Spanish  Research  Agency  (Agencia  Estatal  de  Investigaci\'{o}n) through the grant IFT Centro de Excelencia  Severo  Ochoa  SEV-2016-0597.


\begin{thebibliography}{99}

\bibitem{Zamolodchikov:1986gt}
A.~B.~Zamolodchikov, ``Irreversibility of the Flux of the Renormalization Group in a 2D Field Theory,''
JETP Lett.\  {\bf 43} (1986) 730
[Pisma Zh.\ Eksp.\ Teor.\ Fiz.\  {\bf 43} (1986) 565].


\bibitem{Komargodski:2011xv}
Z.~Komargodski,
``The Constraints of Conformal Symmetry on RG Flows,''
JHEP {\bf 1207} (2012) 069,
arXiv:1112.4538 [hep-th].

\bibitem{Casini:2017vbe}
H.~Casini, E.~Testé and G.~Torroba,
``Markov Property of the Conformal Field Theory Vacuum and the a Theorem,''
Phys.\ Rev.\ Lett.\  {\bf 118} (2017) no.26,  261602,
arXiv:1704.01870[hep-th].

\bibitem{Casini:2012ei}
H.~Casini and M.~Huerta,
``On the RG running of the entanglement entropy of a circle,''
Phys.\ Rev.\ D {\bf 85} (2012) 125016,
arXiv:1202.5650[hep-th].

\bibitem{Solodukhin:2006xv}
S.~N.~Solodukhin,
``Entanglement entropy of black holes and AdS/CFT correspondence,''
Phys. Rev. Lett. \textbf{97} (2006), 201601, 
arXiv:hep-th/0606205 [hep-th].
  
  \bibitem{Kim:2016jwu}
  K.~S.~Kim and C.~Park,
 ``Renormalization group flow of entanglement entropy to thermal entropy,''
  Phys.\ Rev.\ D {\bf 95} (2017) no.10,  106007,
 arXiv:1610.07266[hep-th].
  
    \bibitem{Park:2018ebm}
  C.~Park, D.~Ro and J.~Hun Lee,
  ``c-theorem of the entanglement entropy,''
  arXiv:1806.09072[hep-th].
 
 \bibitem{Casini:2016udt}
  H.~Casini, E.~Teste and G.~Torroba,
  ``Relative entropy and the RG flow,''
  JHEP {\bf 1703} (2017) 089,
  arXiv:1611.00016[hep-th].
  
  
  \bibitem{Solodukhin:2011gn}
  S.~N.~Solodukhin,
``Entanglement entropy of black holes,''
  Living Rev.\ Rel.\  {\bf 14} (2011) 8,
  arXiv: 1104.3712 [hep-th].

\bibitem{Solodukhin:2006ic}
  S.~N.~Solodukhin, ``Entanglement entropy and the Ricci flow,''
  Phys.\ Lett.\ B {\bf 646} (2007) 268,
  arXiv: hep-th/0609045 [hep-th].
  
  \bibitem{Gimre:etal2}
K.~Gimre, C.~Guenther and J.~Isenberg ,
``A geometric introduction to the 2-loop renormalization group flow``.
arXiv:1312.6049v1[math.DG].

\bibitem{Oliynyk:2009rh}
  T.~A.~Oliynyk,
  ``The 2nd order renormalization group flow for non-linear sigma models in 2 dimensions,''
  Class.\ Quant.\ Grav.\  {\bf 26} (2009) 105020.
 arXiv:0904.1241[hep-th].

\bibitem{Branding:2015}
V.~Branding, 
"The normalized second order renormalization group flow on closed surfaces",
arXiv:1503.07462,
  
\bibitem{Perelman:2006un}
G.~Perelman, 
``The Entropy formula for the Ricci flow and its geometric applications,''
arXiv: math/0211159[math-dg].
  
\bibitem{Perelman:2006up}
G.~Perelman,
``Ricci flow with surgery on three-manifolds,''
arXiv: math/0303109[math-dg].


\bibitem{Hamilton:1}
R. ~Hamilton, ``Three-manifolds with positive Ricci curvature''. Journal of Differential Geometry 17 (1982), 255-306. 

\bibitem{Rflow:1}
B.~Chow, D. Knopf. ``The Ricci flow: An introduction. American Mathematical Society. Mathematical Surveys and Monographs, Vol.110 (2004).

\bibitem{Chow:1995}
B. Chow, S.C, Chu. ``A Geometric interpretation of Hamilton's Harnak inequality for the Ricci flow". Mathematical Research Letters 2, 701–718 (1995)



\bibitem{Woolgar:2007vz}
E.~Woolgar, ``Some Applications of Ricci Flow in Physics,''
Can.\ J.\ Phys.\  {\bf 86} (2008) 645.
arXiv: 0708.2144[hep-th].

\bibitem{Carfora:2010iz}
M.~Carfora,``Renormalization Group and the Ricci Flow,''
arXiv:1001.3595[hep-th].

\bibitem{Headrick:2006ti}
M.~Headrick and T.~Wiseman,``Ricci flow and black holes,''
Class.\ Quant.\ Grav.\  {\bf 23} (2006) 6683.
arXiv: hep-th/0606086 [hep-th].

\bibitem{Husain:2008rg}
  V.~Husain and S.~S.~Seahra,
  ``Ricci flows, wormholes and critical phenomena,''
  Class.\ Quant.\ Grav.\  {\bf 25} (2008) 222002
  arXiv:0808.0880 [gr-qc].
  
  \bibitem{LassoAndino:2018zhb}
  Ó.~Lasso Andino,
  ``RG-2 flow, mass and entropy,''
  Class.\ Quant.\ Grav.\  {\bf 36} (2019) 065011,
  arXiv: 1806.10031 [gr-qc].
\bibitem{Gimre:2014jka}
K.~Gimre, C.~Guenther and J.~Isenberg,
``Short-time existence for the second order renormalization group flow in general dimensions,''
Proc.\ Am.\ Math.\ Soc.\  {\bf 143} (2015) no.10,  4397,
arXiv: 1401.1454 [math.DG].

\bibitem{Chow:2006}
B.~Chow, P.~Lu, L.~Ni. ``Hamilton's Ricci flow'' American Mathematical Society, Graduate Studies in Mathematics (77), 2006

\end{thebibliography}
\end{document}